# Gamification, virality and retention in educational online platform. Measurable indicators and market entry strategy

Ilya V. Osipov [1*], Alex A. Volinsky [2*], Vadim V. Grishin [1]

[1] i2istudy.com, Krišjāņa Barona Iela, 130 k-10, Rīga, Lv-1012, Latvija

[2] Department of Mechanical Engineering, University of South Florida, 4202 E. Fowler Ave., ENB118, Tampa FL 33620, USA

[*] Corresponding authors. Email: volinsky@usf.edu; Phone: (813) 974-5658; Fax: (813) 974-3539 (Alex A. Volinsky); Email: ilya@i2istudy.com (Ilya V. Osipov)

**Abstract**

The paper describes gamification, virality and retention in the freemium educational online platform with 40,000 users as an example. Relationships between virality and retention parameters as measurable metrics are calculated and discussed using real examples. Virality and monetization can be both competing and complementary mechanisms for the system growth. The K-growth factor, which combines both virality and retention, is proposed as the metrics of the overall freemium system performance in terms of the user base growth. This approach can be tested using a small number of users to assess the system potential performance. If the K-growth factor is less than one, the product needs further development. If the K-growth factor it is greater than one, the system retains existing and attracts new users, thus a large scale market launch can be successful.

## Introduction

There are numerous products utilizing the freemium model, such as mobile applications, SaaS solutions, software, web applications and others (Seufert, 2014). However, the freemium model is not as simple as it may seem. The authors analyzed the statistics of the users' behavior in the educational collaborative platform available to everybody as shareware and through the freemium model (Osipov, Volinsky & Prasikova, 2014). The platform is a web site for learning foreign languages with users from all over the World. The main idea of the system is based on the fact that regardless of all the grammar learned in college, students are lacking live interactions with the native speakers to increase their spoken language skills (Osipov, Prasikova & Volinsky, 2014). Finding a native speaker is not an easy task, which typically also requires paying for the tutor lessons. The authors noticed, based on the students studying Spanish, that professional teacher is not required to learn basic communication skills. What's needed is a partner, who's ready to help using already prepared materials, a Spanish native speaker. Spanish native speakers are eager to learn English in exchange for teaching Spanish. The authors used the idea of time banking (Marks, 2012; Válek & Jašíková, 2013) to track how much time each user is learning a foreign language and teaching native language.

The authors combined readily available audio-video conferencing technology between the users with the pre-defined lessons, divided into step-by-step cards, understandable by the non-professional teachers, along with the learning/teaching time tracking and an online system of finding and connecting users. This is how the online learning/teaching educational resource, called i2istudy was born (Osipov, 2013). Using this system, which operated from April through August 2014 in the beta mode, and collected over 40,000 users, the authors have conducted several measurements and studies.

Freemium, which is a combination of the words Free and Premium, business model assumes the maximum product market distribution, along with the capture and retention of the largest possible number of users. Part of the users, which for various products varies from 3% to 10%, take advantage of the premium features, allowing the creators not only to pay for the entire system upkeep, including free parts, but also to make a profit (Mäntymäki & Salo, 2011).

Let's consider the basic functions of the freemium products aimed at the mass market:
1. The main (base) function for which the users come (in our case it is learning foreign languages).
2. User retention, including return users.
3. User monetization.
4. User attraction and virality stimulation (existing users attracting new users).

Figure 1 shows a typical cycle of the freemium application. Arrows indicate main user and information flows, including the 4 main cycles:

1. The in-app cycle – the main application cycle, the core cycle, the basic function for which the user decided to use the application (in our case it's learning foreign language with native speaker).
2. The monetization cycle (denoted by the small dollar sign $ in Figure 1). This is an additional cycle, which attracts the most venturous people involved in the process, which represents additional features. This cycle is smaller, since it is not available to all participating users (especially in the freemium business model).
3. The retention cycle is when users leave and subsequently return into the system. To successfully return and retain the users in the system, special means are utilized, from e-mail notifications, social networks and other communication channels reminders of the events, which occurred during the user absence from the system, to gamification.
4. The viral cycle when existing users invite new users from the external environment (e-mail, social networks, blogs, forums, personal websites, applications, and other communication channels), including the new users accommodation.

Besides, the diagram in Figure 1 also shows different user flows into the web application, including organic "word of mouth" users, bookmarks, search engines, motivated and purchased users, along with the invited users (Ellis & Brown, 2014). The downward arrow shows users lost directly from the web front page, as well as from any other of the mentioned cycles. It should be noted that the app cycles: the core, viral, retention and monetization cycles are antagonistic, as they are competing for the user attention, which is always lacking. The system developers must understand which cycles have priority.

When creating freemium products, there are two main business approaches:
1. Purchasing and other paid users (traffic) attraction. Part of the traffic can be monetized by selling additional premium services and attracting new users by spending the money earned. The key factor in this approach is money, thus a successful monetization model is required to involve a significant percentage of users in the pay mechanics to maintain the balance. The positive balance between the revenues from the existing customers minus the cost of attracting new users. Moreover, the cost of attracting new users can be substantial, and there is a risk not to recover this high cost from monetization. Pluses of this approach include fast money earning, and that the K virality factor (K-factor, Reichheld, 2003) can be less than one (discussed below).
2. Involving existing users into the product promotion through virality. It is necessary to ensure that the virality coefficient (the K-factor) is significant, which for a number of products is difficult and even unattainable. The volume of users with this approach is growing exponentially until it reaches saturation (Cohen, 2014). The product can contain features from both approaches, emphasis on monetization and emphasis on expanding the user base. However, given the limited user's attention, one of the approaches must be dominant.

**Viral user base expansion**

Let's consider the second approach, which we call freemium with an emphasis on virality. David Skok, successful venture capitalist, wrote about the freemium virality-emphasized products (Skok, 2009):

"..in a typical business the single biggest expense is sales and marketing, and recognize that offering a free product/service is an extremely smart way to acquire customers at a low cost that can then be monetized in a different way." "Another powerful effect of using the free strategy is that it usually results in a far larger customer base using the free products, who become proponents for your company. This expanded footprint or market share can have a huge effect on the price that acquirers or investors are willing to pay for your company, as they recognize that even though these customers have yet to be monetized, they represent a great potential for future monetization. Twitter and Facebook are two perfect examples of this." "Another way of looking at the importance of footprint or marketshare is to recognize the importance of market leadership. In the tech industry, market leadership is usually self-reinforcing unless the company does stupid things to annoy its customers. Even if you have gained market leadership by giving away a product/service for free, the financial markets and acquirers realize that market leadership is worth a significant premium over niche players that may have more revenue."

However, the strategy of viral user attraction can not be utilized forever. Seufert, the author of the book Freemium Economics (Seufert, 2014) presents a graph, similar to the one shown in Figure 2. All efforts invested in the virality mechanics will not bear fruit when the market niche is already saturated, and all potential users either already use the product, or know about it, but prefer not to use it. Obviously, this is the best time to refocus the product and change the user's attention to monetization, which we discussed as the strategy number 1.

**Virality realization methods (invitations)**

The viral marketing requires several components: the sender, the message and the medium for dissemination, including recipients, along with the context in which the message is received. There are two ways for the user to invite new users:

1. Open invitations - is the viral mechanism, where the user places invitations in social networks, blogs, personal web pages, etc., addressing an undefined set of individuals.

2. Direct personal invitations initiated by the existing users to the potential new users using different means of communication by e-mail, personal communication, social networks, SMS, etc.

Typically it is hard to account for all open invitations. The authors used simplified statistics by calculating how many people were invited by this method, and how many people were able to use this method to initiate invitations. The system calculates how many open invitations were made by each user (via built-in system instruments), and how many new users joined as a result (including open invitations initiated by the user and not generated by the system). Table 1 lists the number of published invitation links (open invitations) and how many users posted these links on the weekly basis.

Statistics reflects only built-in invitation publication mechanism. Table 2 lists the number of the newly joined users invited trough the open links, including all types of invitation. Personal direct invitations allow calculating all parameters and quantifying all steps of the viral cycle. The system accounts for how many users make personal invitations, how many invitations are generated per each user, how many invitations reach the addressee, how many recipients come to the service, and how many register and get involved in the learning/teaching process. Table 3 lists the number of individual invitations and the number of users that sent such invitations, while Table 4 lists the number of users who joined as a result of direct personal invitations.

**The viral cycle and the K-factor**

Let's define the metrics parameters. The user means registered and authorized user of the service.

*dU* stands for daily users;

*dNU* are daily new users;

*dAU* are daily active users (users who spent more than 5 minutes in the system);

*U* is the total number of all users;

*IU* is the number of invited users;

$D_i$ is the total number of invitations per day;

*A$_i$PSU* is the average number of invitations per spreading user (*A$_i$PDSU* is the same per day);

*A$_i$PU* is the average number of invitations per user;

*DIU* is the number of daily invited users;

*IP$_i$* is the ratio of people who accepted an invitation to the number of invitations sent (conversion percentage);

Conceptually the K-factor is the average number of additional users introduced to the product by each user (Seufert, 2014, Chapter 7, p. 170). For practical purposes we calculate the K-factor as the ratio between the users attracted through viral methods in a certain time period, to all active users in this time period (Skok 2009b). Theoretically, we should have used the previous time period, setting it equal to the duration of the viral cycle. However, the length of the viral cycle is difficult to establish, since the reaction to the invitation timeline is very short in our system, as in a typical case, sending invitations and accepting them gets completed in one day. For calculating the K-factor, only new users (*dNU*), or all users in a certain time period (*dU*) can be used, along with the active users in a certain time period (*dAU*). The authors used active users (not considering uninterested users, who spent very little time in the system) as the base, as it gives the most accurate results in our estimates. For practical purposes some sources used only new users (*dNU*) as a base, comparing all users attracted virality with all types of new users. The authors believe this is not quite correct, since all active users contribute to virality and not only new users, thus:

$$Local\ K_{factor} = \frac{dIU}{dAU} \qquad (1)$$

The term K-factor comes from epidemiology, "in which a virus having a K-factor of 1 is in a "steady" state of neither growth nor decline, while a K-factor greater than 1 indicates exponential growth and a K-factor less than 1 indicates exponential decline." (Lee, 2008). The K-factor, which is also called viral coefficient in the literature, can be calculated as the number of invitations sent by each user multiplied by the conversion percentage of the new users (Fong, 2014). For example, if the average number of invitations per user is 5, and 20% of the invitees register in the system, i.e. become new users, then the K-factor = 5*0.2=1. Time-independent K-factor averaged over the whole time of the system operation the authors call the global K-factor, which is calculated in the following way. The conversion percentage, *IPi* is the number of invited people, *IU*, divided by the number of invitations, *i*:

$$IP_i = \frac{IU}{i} \qquad (2)$$

*AiPU* is the average number of invitations per user, calculated as the average number of invitations per user, *i*, divided by the total number of users, *U*:

$$A_i PU = \frac{i}{U} \quad (3)$$

Then the global K-factor is calculated as the product of the average number of invitations per user, *AiPU* (equation 3) and the conversion percentage, *IPi* (equation 2):

$$Global\ K_{factor} = A_i PU \cdot IP_i \quad (4)$$

The K-factor dynamics reflects the users' mood swings, and how they react to the introduction, activation or deactivation of one or another viral mechanics, involving them in the activities of inviting new users, and whether these mechanics are well accepted. Thus for practical purposes the authors utilized the local K-factor, calculated daily. The authors call it the daily K-factor, *dK*-factor, which is based on the active audience, *dAU*. This daily K-factor is considered the most important parameter of the viral cycle, which was used in the construction of viral mechanisms of the project:

$$Daily\ K_{factor}, dK = dAU \cdot IP_i \quad (5)$$

Figure 3 shows the i2istudy foreign language educational platform K-factor dynamics.

It is important to understand that if the K-factor is less than unity (e.g. 50%), in the absence of retention (when the loyalty of existing users is zero), the system growth attenuates. In the best scenario, such virality mechanism partially compensates the users' loss as a result of the normal loyalty retention cycle decrease. This K-factor increases the effectiveness of paid user attraction. For example, purchased 100 paying users get involved in the viral mechanics and invite additional 50 users, which reduced the average price of each user and saved the budget. If the K-factor is greater than 1 (say it is 200%), it leads to the geometric progression growth of the user base. For example, purchased 100 users attract 200 new people, and if the K-factor remains the same, the new users will attract 400 new people, and so on. Virality works as long as the entire mass of potential users will not reach saturation in their social matrix and a given market.

**Virality, retention and monetization relationship**

In his book Freemium Economics, Eric Benjamin Seufert (Seufert, 2014) in the Virality and Retention section, on page 175 said: "Virality and retention exist on opposite sides of the acquisition threshold: virality describes how users are introduced to a product, and retention describes how long users remain with a product. But in essence, both sets of metrics measure the same general sense of delight users feel for a product, manifested in different ways. To that end, virality and retention generally exhibit a positively correlated relationship: products that users are inclined to return to over a long period of time are also likely to be products that users invite others to join."

In our opinion, virality and retention are characteristics amenable to manipulation by the creators of the product. Even a weak product can successfully maintain good retention and virality performance, if appropriate mechanics and effects (impact, gamification) are well integrated into the product and successfully motivate users to these actions. This situation resembles a grocery store, where buyers are manipulated by the layout, marketing, branding, packaging and a discount system, and buy groceries that are not the best and healthy as a result (Glanz, Bader & Iyer, 2012).

Certainly all three parameters: virality, retention and monetization are related. Users with high product loyalty get increasingly involved in the mechanics of virality and monetization (Fields & Cotton, 2012; Eyal & Hoover (2014). Despite competition for the user's attention, these mechanisms may spur one another, and all sorts of techniques, such as gamification, which is usually considered in the literature as part of retention, can serve monetization and virality. Oddly enough, monetization, can also spur virality and retention. For example, premium paid services can be alternatively earned by participation in the viral and gamification programs. The dollar price of these premium options demonstrates their value to the users. For example, when the user knows the cost of acquiring new premium options for real money, it may be easier to motivate the user to earn these premium options by performing certain tasks and actions, such as inviting friends.

It is important that the experience of using the main basic functions of the product cause admiration, then the virality and retention mechanics come into play. If they are unbalanced, for example, with perfect virality and poor retention, the growth of the user base, caused by the successful virality, will compensate for the loss of the same base due to disloyal users. The opposite situation of poor virality with excellent retention leads to the product and its user base stagnation, and eventual defeat by the competitors.

Coefficient of the product audience growth, K-growth, can be expressed as a sum of the coefficients of the viral K-factor and the retention factor, K-retention:

$$K_{growth} = K_{factor} + K_{retention} \qquad (6)$$

Equation 6 is the main formula of the freemium product growth, based on the viral spreading. It is clear that this formula does not take into account alternative methods of attracting users, such as paid users and organic users, who came through search engines, word of mouth, or due to the brand popularity. K-retention is always less than one over a long period of time, since no products can retain its audience 100% at all times.

$$Local\ K_{retention} = \frac{dU - dNU}{dU_{-1}} \quad (7),$$

where *dU* is the daily audience for a given day; *dU₋₁* is the previous day audience and *dNU* are the new users for this time period. It's convenient to use only active audience for calculations, by taking into account only the new users that have become active, but not all registered users. Similar situation is with the new invited users, among which only active users are accounted for:

$$Local\ K_{retention_A} = \frac{dAU - dNU}{dAU_{-1}} \quad (8)$$

For example, if the viral K-factor is 20 %, and K-retention factor is 90% (i.e. 9 out of 10 people are coming the next day), the growth coefficient will be 0.2 + 0.9 = 1.1 and the system will grow on its own by 10% of its daily (or other accounting period) audience. Coefficient of the system self growth can be represented as:

$$K_{growth} = \frac{dAU_1 - dNU + dIU}{dAU_0} \quad (9)$$

This is the ratio of the audience from the next period of time without accounting for the new users, but accounting for the users invited by the viral techniques, divided by audience from the previous time period. If *dNU* and *dIU* are equal, then all new users get involved through viral methods exclusively (there is no paid and organic traffic), then in this case:

$$K_{growth} = \frac{dAU}{dAU_{-1}} \quad (10)$$

Table 5 shows the data and the corresponding parameters calculations for the i2istudy product.

As a side note, it is necessary to take into account that the very properties of the product may be a barrier to its viral spread. For example, users absolutely don't want to advertise to their friends that they participate in dating services (Blackhart, Fitzpatrick & Williamson, 2014), which negates any virality efforts. On the contrary, the users promote their morning runs and other physical exercises, even without strong viral mechanics and ingenious motivations (Kamal, Fels & Fergusson, 2014; Loss,

Lindacher, Curbach, 2014). An obstacle to the viral spread may be excessive annoyance of the viral mechanics, which can be negatively perceived by the existing users, and even considered as spam by the invitation recipients (Grimes, Hough & Signorella, 2007). In addition, it's a common mistake to promote business-to-business (b2b) services using virality methods, which usually gives poor results (Davidson, 2014), with the exception of individual cases (Grindeanu, 2014).

If the K-growth factor is less than one, then the product can not grow and looses users with all the consequences for the product and its team. However, if the team can achieve the K-growth factor greater than one, the product grows exponentially. It is the ultimate goal for the product team to achieve non-paid user base growth. This is necessary to achieve the project's capitalization exceeding the investment in the purchase of the user base. Note that the positive K-growth factor can compensate for other shortcomings, such as the quality of the product itself. For the overall project development strategy, investment in virality and retention is a viable alternative to investments in advertising and public relations. This is often a cost-effective solution, since compared with the cost of development (programmers salaries, etc.), marketing and associated staff costs can be quite high. The feemium product development diagram in terms of the K-growth factor is shown in Figure 4.

**Conclusions**

When building a freemium product, it is wise to take it to the market and to work out the viral and retention mechanics on small volumes of paid audience, since these mechanics can be easily evaluated statistically and analyzed. Having a positive K-growth factor, venture capital funds can be attracted, and the project can be brought to a large market (Figure 4). As for our specific product, which was used to conduct these studies, it is clear that the value of the K-growth factor varies around 40% (last line in Table 5), which is not a satisfactory. As a result, the product is currently being reworked and improved.

**Acknowledgements**

The authors would like to thank the i2istudy.com team for their dedicated efforts: Anna Prasikova, Ilya Poletaev, Andrei Poltanov, Elena Bogdanova, Vildan Garifulin, Mihail Denisov and Franziska Rinke.

Language Learning.

**Table Captions**

Table 1. The number of published invitation links (open invitations) and how many users post these links on the weekly basis.

Table 2. The number of the newly joined users invited trough the open links, including all types of invitations.

Table 3. The number of individual invitations and the number of users that sent these invitations.

Table 4. The number of users who joined as a result of direct personal invitations.

Table 5. Data and parameters calculations for the i2istudy product.

**Figure Captions**

Figure 1. Typical cycles of the freemium application.

Figure 2. Saturation point in the total number of users with time. Adapted from Seufert, 2014.

Figure 3. Weekly K-factor dynamics for the i2istudy project.

Figure 4. Feemium product development diagram in terms of the K-growth factor.

Table 1. The number of published invitation links (open invitations) and how many users post these links on the weekly basis.

| week | 12.05 → 18.05 | 19.05 → 25.05 | 26.05 → 01.06 | 02.06 → 08.06 | 09.06 → 15.06 | 16.06 → 22.06 | 23.06 → 29.06 | 30.06 → 06.07 | 07.07 → 13.07 | 14.07 → 20.07 | 21.07 → 27.07 | 28.07 → 03.08 | 04.08 → 10.08 | 11.08 → 17.08 | 18.08 → 24.08 |
|---|---|---|---|---|---|---|---|---|---|---|---|---|---|---|---|
| Published invitation links | | | 10 | 17 | 7 | 19 | 23 | 12 | 15 | 7 | 7 | 12 | 33 | 14 | 15 |
| Number of users publishing links | | | 8 | 8 | 3 | 10 | 11 | 9 | 10 | 5 | 5 | 10 | 28 | 13 | 10 |

Table 2. The number of the newly joined users invited trough the open links, including all types of invitations.

| week | 21.04 → 27.04 | 28.04 → 04.05 | 05.05 → 11.05 | 12.05 → 18.05 | 19.05 → 25.05 | 26.05 → 01.06 | 02.06 → 08.06 | 09.06 → 15.06 | 16.06 → 22.06 | 23.06 → 29.06 | 30.06 → 06.07 | 07.07 → 13.07 | 14.07 → 20.07 | 21.07 → 27.07 | 28.07 → 03.08 | 04.08 → 10.08 | 11.08 → 17.08 | 18.08 → 24.08 |
|---|---|---|---|---|---|---|---|---|---|---|---|---|---|---|---|---|---|---|
| Invited by public link | | 21 | 15 | 7 | 11 | 13 | 47 | 81 | 70 | 10 | 26 | 29 | 25 | 28 | 30 | 35 | 40 | 15 |

Table 3. The number of individual invitations and the number of users that sent these invitations.

| week | 21.04 → 27.04 | 28.04 → 04.05 | 05.05 → 11.05 | 12.05 → 18.05 | 19.05 → 25.05 | 26.05 → 01.06 | 02.06 → 08.06 | 09.06 → 15.06 | 16.06 → 22.06 | 23.06 → 29.06 | 30.06 → 06.07 | 07.07 → 13.07 | 14.07 → 20.07 | 21.07 → 27.07 | 28.07 → 03.08 | 04.08 → 10.08 | 11.08 → 17.08 | 18.08 → 24.08 |
|---|---|---|---|---|---|---|---|---|---|---|---|---|---|---|---|---|---|---|
| Requests total | 44 | 156 | 212 | 32 | 618 | 438 | 942 | 595 | 1494 | 838 | 1073 | 3029 | 5823 | 2693 | 2732 | 3156 | 1129 | 1683 |
| Requested users total | 2 | 22 | 14 | 4 | 17 | 37 | 41 | 36 | 90 | 44 | 80 | 127 | 103 | 73 | 52 | 83 | 40 | 24 |

Table 4. The number of users who joined as a result of direct personal invitations.

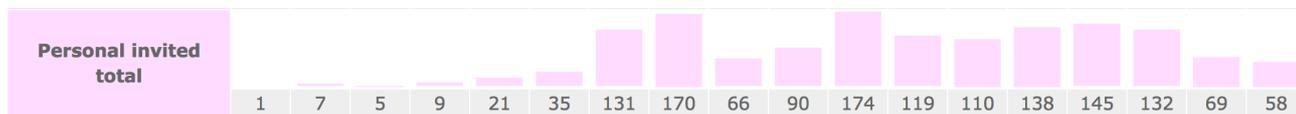

Table 5. Data and parameters calculations for the i2istudy product.

| week | 05.05 → 11.05 | 12.05 → 18.05 | 19.05 → 25.05 | 26.05 → 01.06 | 02.06 → 08.06 | 09.06 → 15.06 | 16.06 → 22.06 | 23.06 → 29.06 | 30.06 → 06.07 | 07.07 → 13.07 | 14.07 → 20.07 | 21.07 → 27.07 | 28.07 → 03.08 | 04.08 → 10.08 | 11.08 → 17.08 | 18.08 → 24.08 |
|---|---|---|---|---|---|---|---|---|---|---|---|---|---|---|---|---|
| All active users (xAU) | 297 | 801 | 867 | 979 | 1080 | 1213 | 1827 | 2126 | 1763 | 2624 | 2572 | 1924 | 1716 | 1810 | 1576 | 947 |
| New active users (xNU) | 239 | 575 | 572 | 643 | 695 | 776 | 1358 | 1464 | 1140 | 1881 | 1643 | 1129 | 998 | 1066 | 824 | 358 |
| Invited active users (xIU) | 5 | 9 | 10 | 32 | 54 | 81 | 85 | 41 | 73 | 68 | 52 | 65 | 77 | 78 | 52 | 42 |
| K-Factor = xIU / xAU, percent | 2 | 1 | 1 | 3 | 5 | 7 | 5 | 2 | 4 | 3 | 2 | 3 | 4 | 4 | 3 | 4 |
| K-Retention = (xAU - xNU) / xAU$_{-1}$, percent | | 76 | 37 | 39 | 39 | 40 | 39 | 36 | 29 | 42 | 35 | 31 | 37 | 43 | 42 | 37 |
| K-Growth = (xAU - xNU + xIU) / xAU$_{-1}$, percent | | 79 | 38 | 42 | 45 | 48 | 46 | 38 | 33 | 46 | 37 | 33 | 41 | 48 | 44 | 40 |

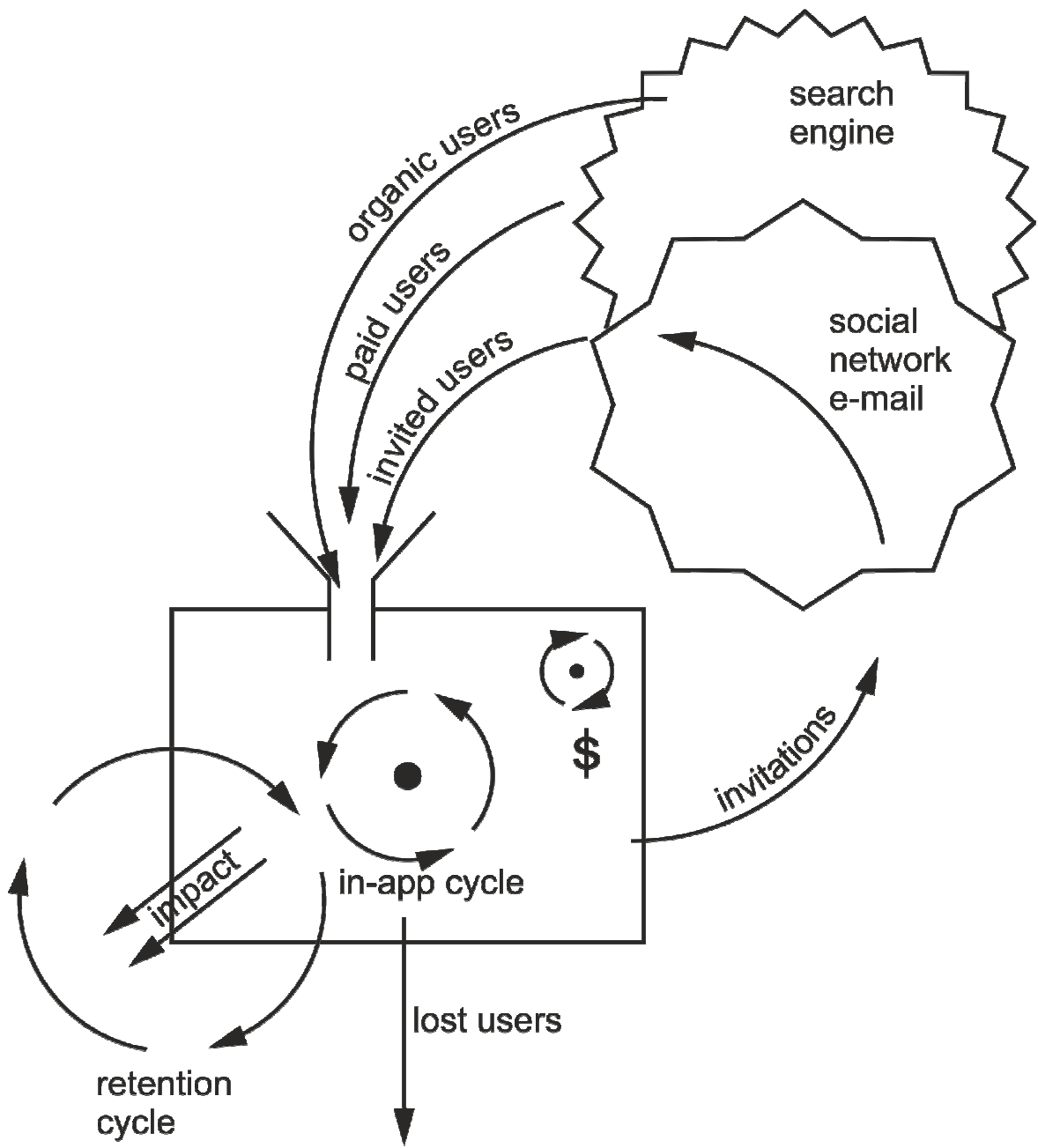

Figure 1. Typical cycles of the freemium application.

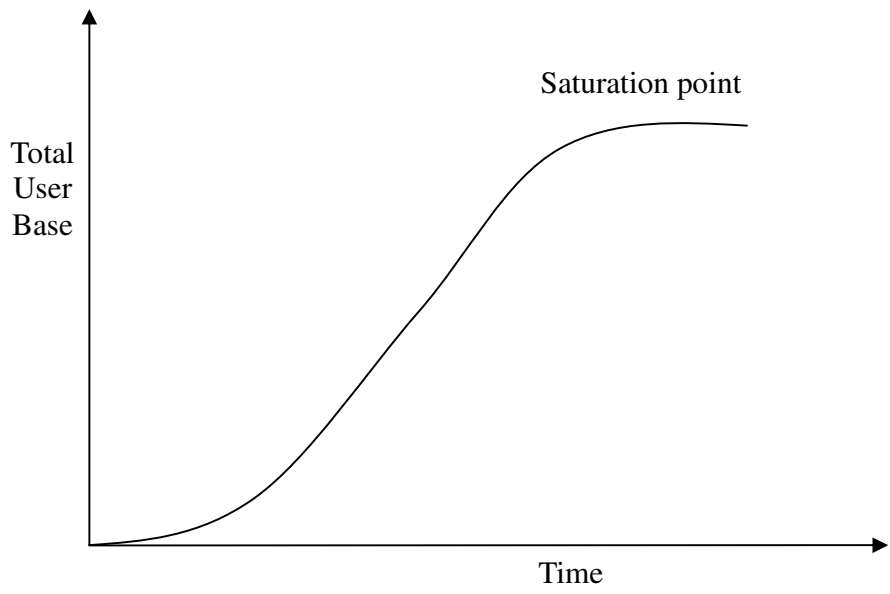

Figure 2. Saturation point in the total number of users with time. Adapted from Seufert, 2014.

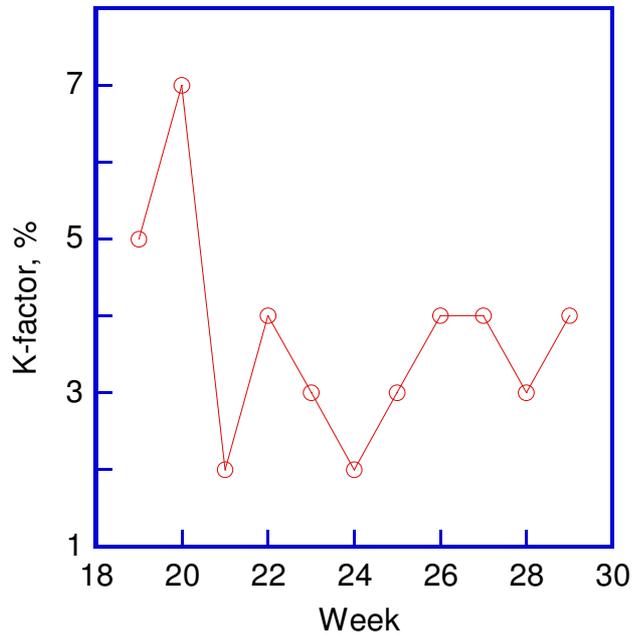

Figure 3. Weekly K-factor dynamics for the i2istudy project.

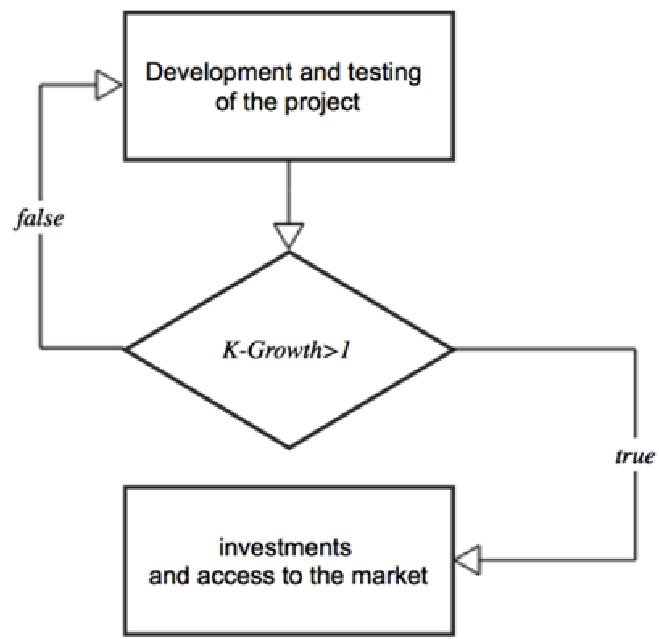

Figure 4. Feemium product development diagram in terms of the K-growth factor.